\documentclass[reprint, showpacs, superscriptaddress, aps, prfluids]{revtex4-2}

\usepackage{amsmath}
\usepackage{graphicx}
\usepackage{epstopdf}
\usepackage{color}
\usepackage{amssymb}
\usepackage{amsmath}
\usepackage{tabularx}
\usepackage{bm}
\usepackage{hyperref}
\usepackage{ltxtable}
\usepackage{natbib}
\usepackage{longtable}
\usepackage[normalem]{ulem}
\usepackage{float}
\usepackage[dvipsnames,table,xcdraw]{xcolor}
\usepackage{cleveref}
\usepackage{multirow}

\newcommand{\la}{\left\langle}
\newcommand{\ra}{\right\rangle}
\newcommand{\be}{\begin{equation}}
\newcommand{\ee}{\end{equation}}
\newcommand{\bse}{\begin{subequations}}
\newcommand{\ese}{\end{subequations}}
\newcommand{\bea}{\begin{eqnarray}}
\newcommand{\eea}{\end{eqnarray}}
\newcommand{\ba}{\begin{array}}
\newcommand{\ea}{\end{array}}

\hypersetup{
colorlinks,
citecolor=blue,
filecolor=blue,
linkcolor=blue,
urlcolor=blue}

\begin{document}

\title{Kolmogorov Scaling for Total Energy and Cross Helicity in Magnetohydrodynamic Turbulence}
\author{Manthan Verma}
\email{manver@iitk.ac.in}
\affiliation{Department of Physics, Indian Institute of Technology Kanpur, Kanpur 208016, India}
\author{Abhishek K. Jha}
\email{abhijha@iitk.ac.in}
\affiliation{Department of Physics, Indian Institute of Technology Kanpur, Kanpur 208016, India}
\author{Mahendra K. Verma}
\email{mkv@iitk.ac.in}
\affiliation{Department of Physics, Indian Institute of Technology Kanpur, Kanpur 208016, India}
\affiliation{Kotak School of Sustainability, Indian Institute of Technology, Kanpur 208016, India}

\date{\today}

\begin{abstract}
The problem of scaling in isotropic magnetohydrodynamic (MHD) turbulence has remained unresolved, with competing predictions of $k^{-5/3}$ (Kolmogorov) and $k^{-3/2}$ (Iroshnikov-Kraichnan) scalings. In this paper, we  address this long-standing controversy using high-resolution numerical simulations on $8192^2$ and $1536^3$ grids. We show that the total energy and cross helicity spectra are closer to $k^{-5/3}$ than $k^{-3/2}$. The fluxes and structure functions of the total energy and cross helicity also demonstrate robust support for Kolmogorov scaling. The magnetic energy shows $k^{-5/3}$ spectrum, but the kinetic energy exhibits  $k^{-3/2}$ spectrum; the latter spectrum is  due to the energy transfers from the magnetic field to the velocity field.

\end{abstract}
\maketitle

\section{Introduction}

Many astrophysical systems, such as the solar corona,  solar convection zone, and galaxies, contain turbulent quasineutral plasma that is described using magnetohydrodynamics (MHD)~\cite{Biskamp:book:MHDTurbulence}. Hence, MHD turbulence is of major interest to physicists. However, MHD turbulence is quite complex due to the interplay of velocity (\textbf{u}) and magnetic (\textbf{b}) fields and numerous nonlinear terms.  Researchers have proposed spectral theories with varying predictions, e.g., with energy spectra---$k^{-5/3}$, $k^{-3/2}$, $k^{-2}$~\cite{Biskamp:book:MHDTurbulence,Verma:PR2004,Schekochihin:JPP2022,Zhou:PR2021,Sagaut:book,Miesch:SSR2015}.   In this paper,  we analyse the isotropic energy spectra, energy fluxes, and structure function using numerical simulations and demonstrate that MHD turbulence follows Kolmogorov-like turbulence phenomenology, which yields $k^{-5/3}$ spectrum.

Kolmogorov~\cite{Kolmogorov:DANS1941Dissipation,Kolmogorov:DANS1941Structure} showed that for a steady, homogeneous, and isotropic turbulent flow, under the limit of kinematic viscosity $\nu \rightarrow 0$, the third-order structure function $S^u_3(l) = \la [\{ {\bf u(x+l)-u(x)} \} \cdot \hat{\bf l}]^3 \ra = -(4/5) \epsilon_u l$, where the velocity field is measured at two positions, \textbf{x}  and \textbf{x+l}, and $\epsilon_u$  is the dissipation rate of the kinetic energy. The above law also leads to the energy spectrum for the velocity field as $E_u(k) = K_\mathrm{Ko} \epsilon_u^{2/3} k^{-5/3}$, where $K_\mathrm{Ko} $ is Kolmogorov's constant.  Kolmogorov's theory is often a starting point for modeling complex turbulent flows, including MHD turbulence.  

In MHD turbulence, two kinds of Alfv\'{e}n waves propagate in  directions parallel and antiparallel to  the mean magnetic field (${\bf B}_0$) or large-scale magnetic field~\cite{Biskamp:book:MHDTurbulence}.  \citet{Kraichnan:PF1965MHD} and \citet{Iroshnikov:SA1964} argued that these forward and backward moving waves interact for a short duration [$(k B_0)^{-1}$], leading  to  the  following kinetic energy spectrum [$E_u(k)$] and magnetic energy spectrum [$E_b(k)$]:
\be 
E_u(k) \approx E_b(k) \approx K_\mathrm{IK} (\epsilon^T B_0)^{1/2} k^{-3/2},
\label{eq:IK}
\ee
where $\epsilon^T$ is the dissipation rate of the total energy (kinetic+magnetic);  $ B_0$ is the strength of the mean magnetic field or the large-scale magnetic field; and $K_\mathrm{IK}$ is a constant. In this framework, referred to as Iroshnikov-Kraichnan or  \textit{IK scaling},  \citet{Dobrowolny:PRL1980} showed that the inertial-range energy fluxes [$\Pi^\pm$(k)] of the Els\"{a}sser variables  ${\bf z}^\pm =  {\bf u \pm b}$ are equal, i.e., $\Pi^+(k) = \Pi^-(k)$ for any $E^+(k) / E^-(k)$ ratio. This regime where the linear terms (${\bf B}_0 \cdot \nabla) {\bf z}^\pm$  dominates the nonlinear terms $({\bf z ^\mp \cdot  \nabla) z^\pm}$ is called \textit{weak turbulence}~\cite{Kraichnan:PF1965MHD,Biskamp:book:MHDTurbulence,Galtier:JPP2000,Verma:PR2004,Schekochihin:JPP2022}. 

Interestingly, the above $k^{-3/2}$ energy spectrum has a strong resemblance to Kraichnan's earlier prediction on hydrodynamic turbulence. 
Using the Direct Interaction Approximation (DIA), \citet{Kraichnan:JFM1959} proposed that the energy spectrum of hydrodynamic turbulence scales as $E_u(k) \approx  (\epsilon_u U_0)^{1/2} k^{-3/2}$, where $U_0$ is the rms velocity, and $\epsilon_u$   is the energy  dissipation rate. Unfortunately, the above prediction is not borne in experiments and numerical simulations.  The contradiction is attributed to the  lack of ``random Galilean invariance" and ``sweeping effect" in hydrodynamic turbulence~\cite{Kraichnan:PF1964Eulerian}. It is worth noting, however, that  the Alfv\'{e}n velocity cannot be eliminated in MHD turbulence through Galilean invariance, making the  $k^{-3/2}$ scaling as a viable possibility.

For \textit{strong turbulence}, where the nonlinear terms dominate the linear term, \citet{Marsch:RMA1991} proposed the following isotropic \textit{Kolmogorov scaling}~(also see \cite{Goldstein:ARAA1995,Lithwick:ApJ2007,Beresnyak:ApJ2008a,Verma:PR2004}):
\be
E^\pm(k) = K^\pm \frac{(\Pi^\pm)^{4/3}}{(\Pi^\mp)^{2/3}} k^{-5/3},
\label{eq:Kolm_pm}
\ee
where $K^\pm$ are  the  constants of $O(1)$, similar to Kolmogorov's constant for hydrodynamic turbulence;  and $\Pi^\pm$ are the inertial-range energy fluxes for ${\bf z}^\pm$, respectively. Note that $\Pi^\pm = \epsilon^\pm $, the dissipation rates of ${\bf z}^\pm$ in the steady state. In this framework, $\Pi^+(k) > \Pi^-(k)$ when $E^+(k) > E^-(k)$, and vice versa, which is unlike IK and \citet{Dobrowolny:PRL1980}'s predictions that $\Pi^+(k) = \Pi^-(k)$ irrespective of $E^+(k) / E^-(k)$ ratio. The flows with  $E^+  \ne E^-$, which are referred to as  \textit{Alfv\'{e}nic} or \textit{imbalanced MHD}~\cite{Goldstein:ARAA1995,Lithwick:ApJ2007,Beresnyak:ApJ2008a}, are quantified using cross helicity, $\int d{\bf r}({\bf u \cdot b})/2$, and normalized cross helicity, $\sigma_c = (E^+-E^-)/(E^++E^-) = \int d{\bf r} 2({\bf u \cdot b})/\int d{\bf r} (u^2+b^2)$~\cite{Goldstein:ARAA1995}. \citet{Zhou:JGR1990model} and \citet{Zhou:RMP2004} constructed a combined model that yields dual energy spectra: $k^{-5/3}$ at small wavenumbers and $k^{-3/2}$ at large wavenumbers.

In this paper, we show that the spectra of the total energy [$(u^2+b^2)/2$ or $\{(z^+)^2 + (z^-)^2)\}/4$] and cross helicity  [${\bf u \cdot b}/2$ or $\{ (z^+)^2 - (z^-)^2)\}/8$] follow Kolmogorov scaling.  In particular, the spectrum of the total energy scales as
\be
E_T(k) = K_\mathrm{KoM}  (\Pi_T)^{2/3} k^{-5/3},
\label{eq:Kolm}
\ee
where $\Pi_T = (\Pi^++\Pi^-)/2$ is the inertial-range  flux of the total energy, and $K_\mathrm{KoM}$ is Kolmogorov's constant for MHD turbulence. Here, $K_\mathrm{KoM} =  [K^+ + K^-/\zeta^2] \zeta^{4/3} (1+\zeta)^{-2/3} 2^{-1/3}$ with $\zeta =\epsilon^+/\epsilon^-$. Note that the inertial-range cross helicity flux, $\Pi_{H_c} = (\Pi^+-\Pi^-)/4$, is positive when $E^+(k)/E^-(k)>1$ and negative when $E^+(k)/E^-(k)<1$.  The inertial-range fluxes $\Pi_T$ and  $\Pi_{H_c}$ match with the corresponding injection rates $\epsilon^T_\mathrm{inj}$ and $ \epsilon^{H_c}_\mathrm{inj}$, respectively. As we demonstrate later, our numerical results disagree with the IK scaling where $E_T(k)\sim k^{-3/2}$ and  $H_c(k) \approx 0$  because $\Pi^+(k) \approx \Pi^-(k)$ \cite{Lithwick:ApJ2003}.  

We also show that the kinetic energy spectrum $E_u(k) \sim k^{-3/2}$ and the magnetic energy spectrum  $E_b(k) \sim k^{-5/3}$, similar to some of the past results ~\cite{Alexakis:PRL2013,Podesta:ApJ2007,Jiang:JFM2023}.  We attribute these divergent spectral indices to the energy transfer from the magnetic field to the velocity field. In contrast, the total energy and cross helicity show definitive $k^{-5/3}$ spectra because their fluxes are constant in the inertial range.  In the same vein, the companion paper~\cite{Verma:MHD_PRF} shows Kolmogorov scaling for the Els\"{a}sser variables, with their inertial-range fluxes being constant.



Now we consider anisotropic MHD turbulence with finite ${\bf B}_0$. \citet{Iroshnikov:SA1964} and \citet{Kraichnan:PF1965MHD} projected Eq.~(\ref{eq:IK}) to hold for a large ${\bf B}_0$. However, for ${\bf B}_0 \ne 0$, \citet{Sridhar:ApJ1994} objected to the IK scaling by arguing that the three-wave resonant interactions are absent in MHD turbulence. But, \citet{Galtier:JPP2000} showed that three-wave interactions are present in MHD turbulence and predicted that $E(k_\perp) \sim k_\perp^{-2}$, where $k_\perp$ and $k_\parallel$ are, respectively, the perpendicular and parallel components of \textbf{k} in relation to $\textbf{B}_0$. For the strong turbulence regime with moderate ${\bf B}_0$, \citet{Goldreich:ApJ1995} proposed that $k_\parallel B_0 \approx k_\perp z_{k_\perp}$ that yields $E^\pm(k_\perp) \sim k_\perp^{-5/3}$; \citet{Lithwick:ApJ2007} generalized this framework to nonzero $\sigma_c$ and obtained Eq.~(\ref{eq:Kolm}), with one change, which is $k \rightarrow k_\perp$.   Later, \citet{Boldyrev:PRL2006} argued in favour of $k_\perp^{-3/2}$ energy spectrum based on dynamic alignment between \textbf{u} and \textbf{b} fields. For anisotropic MHD turbulence, the numerical simulations of \citet{Cho:ApJ2000} and \citet{Beresnyak:PRL2011} support $-5/3$ spectral exponent, but those of \citet{Mason:PRE2008} support $-3/2$ spectral exponent.

For   \textit{isotropic} and  \textit{anisotropic} MHD turbulence, numerical studies  present diverse picture: some support Kolmogorov scaling \cite{Muller:PRL2000,Biskamp:PP2000,Beresnyak:PRL2011}, some others point to IK scaling \cite{Biskamp:PFB1989,Biskamp:PP2001,Mason:PRL2006}, and some remain inconclusive \cite{Debliquy:PP2005,Sahoo:NJP2011}.  M\"{u}ller and Biskamp~\cite{Muller:PRL2000,Biskamp:PP2000} report $k^{-5/3}$ spectra for three dimensions (3D), but \citet{Biskamp:PFB1989} and \citet{Biskamp:PP2001} report $k^{-3/2}$ spectra for two dimensions (2D).  Unfortunately, ambiguity in the field persists because the spectral indices $-5/3$ and $-3/2$  are quite close.  Furthermore, $E_u(k)$ and $E_b(k)$ yield different spectral exponents~~\cite{Podesta:ApJ2007,Alexakis:PRL2013,Jiang:JFM2023}.   In this paper we show that the \textit{isotropic} spectra, fluxes, and structure functions of the total energy and cross helicity are in agreement with Kolmogorov scaling, rather than  IK scaling. We also show that  the  spectra and the flux of total energy also support Kolmogorov scaling for the anisotropic MHD turbulence. In future work we will explore a more detailed analysis of anisotropic case with higher order of anisotropy.


\section{Energy fluxes in MHD turbulence}

The equations for incompressible MHD are~\cite{Biskamp:book:MHDTurbulence}
\bea
\frac{\partial \mathbf{u}}{\partial t} + (\mathbf{u} \cdot \mathbf{\nabla}) \mathbf{u} &=& -\nabla p + (\mathbf{B} \cdot \mathbf{\nabla}) \mathbf{B} + \mathbf{F_u} + \nu \nabla^2 \mathbf{u},
\label{eq:navier_stokes} \\
\frac{\partial \mathbf{B}}{\partial t} + (\mathbf{u} \cdot \mathbf{\nabla}) \mathbf{B} &=& (\mathbf{B} \cdot \mathbf{\nabla}) \mathbf{u} + \mathbf{F_b} + \eta \nabla^2 \mathbf{B},
\label{eq:MHD} \\
\mathbf{\nabla} \cdot \mathbf{u} &=& 0,
\label{eq:velocity_divergence} \\
\mathbf{\nabla} \cdot \mathbf{B} &=& 0,
\label{eq:magnetic_divergence}
\eea
where $\mathbf{u},\mathbf{B}, p$ are the velocity, magnetic, and pressure (magnetic+thermodynamic) fields, respectively; $\mathbf{B = B_0 + b}$, where $\mathbf{B_0}$ and \textbf{b} are, respectively, the mean magnetic field and its  fluctuations; $\nu$ and $\eta$ are the kinematic viscosity and magnetic diffusivity, respectively; $\mathbf{F}_u$ and $\mathbf{F}_b$ are the external forces applied to the velocity and magnetic fields, respectively. We force both \textbf{u} and \textbf{b} fields to maintain a steady total energy and cross helicity, while maintaining negligible cross helicity. Refer to \cite{Carati:JoT2006,Teimurazov:JAMTP2018} and the companion paper \cite{Verma:MHD_PRF} for further details on the forcing scheme. In this paper we mainly investigate isotropic MHD turbulence (${\bf B}_0 =0$), along with several anisotropic cases with ${\bf B}_0 = \hat{z}$ and $3\hat{z}$. In the inviscid limit, the kinetic energy ($\int d{\bf r} u^2/2$) and the magnetic energy ($\int d{\bf r} b^2/2$) are not individually conserved, but the total energy ($T = \int d{\bf r} [u^2 +b^2]/2)$ and the total cross helicity ($H_c = \int d{\bf r} [{\bf u \cdot b}]/2$) are conserved. In this paper, we focus on the total energy and the total cross helicity. 

In the companion paper \cite{Verma:MHD_PRF}, we reported the  spectra, fluxes, and the structure functions for the Els\"{a}sser variables (${\bf z}^\pm$). In this paper, we will compute these quantities for the velocity and magnetic fields, as well as for the total energy and cross helicity. For \textbf{u} and \textbf{b}, the nonlinear terms, $(\mathbf{u} \cdot \mathbf{\nabla}) \mathbf{u}$, $(\mathbf{b} \cdot \mathbf{\nabla}) \mathbf{b}$, $(\mathbf{u} \cdot \mathbf{\nabla}) \mathbf{b}$, and $(\mathbf{b} \cdot \mathbf{\nabla}) \mathbf{u}$, yield multiscale energy transfers, which are conveniently described using Fourier transforms~\cite{Biskamp:book:MHDTurbulence,Verma:PR2004,Verma:book:ET}. The modal energies for $\bf u$ and $\bf b$ are $E_u({\bf k}) = |{\bf u}({\bf k})|^2/2$ and $E_b({\bf k}) = |{\bf b}({\bf k})|^2/2$, respectively, whose dynamical equations are~\cite{Verma:PR2004,Verma:book:ET}
\bea
\frac{d}{dt}E_u(\mathbf{k}) &= &-\sum_p \Im[\{\mathbf{k \cdot u(q)\}\{u(p)\cdot u^{*}(k)\}}] \notag \\
& &+ \sum_p \Im[\{\mathbf{k \cdot B(q)\}\{b(p)\cdot u^{*}(k)\}}]  \notag \\
& & + \sum_{\mathbf{k}} \mathrm{Re}\left[\mathbf{F}_{u}(\mathbf{k}) \cdot \mathbf{u}^{*}(\mathbf{k})\right] - 2 \nu k^2  E_u(\mathbf{k}) \nonumber \\
&= & T^{uu}(\mathbf{k}) + T^{ub}(\mathbf{k}) + \mathcal{F}^{u}(\mathbf{k}) - D^{u}(\mathbf{k}),
\label{eq:Ek_u} \\
\frac{d}{dt}E_b(\mathbf{k}) &= & -\sum_p \Im[\{\mathbf{k \cdot u(q)\}\{b(p)\cdot b^{*}(k)\}}]  \notag \\
&& + \sum_p \Im[\{\mathbf{k \cdot B(q)\}\{u(p)\cdot b^{*}(k)\}}]  \notag \\
&& + \sum_{\mathbf{k}} \mathrm{Re}\left[\mathbf{F}_{b}(\mathbf{k}) \cdot \mathbf{b}^{*}(\mathbf{k})\right] - 2\eta k^2  E_b(\mathbf{k}) \nonumber \\
&= &T^{bb}(\mathbf{k}) + T^{bu}(\mathbf{k}) + \mathcal{F}^{b}(\mathbf{k}) - D^{b}(\mathbf{k}),
\label{eq:Ek_b}
\eea
where $\mathbf{k} = \mathbf{p} + \mathbf{q}$; $\mathcal{F}^{u}(\bold{k})$ and $\mathcal{F}^{b}(\bold{k})$ are the respective energy injection rates to  $\mathbf{u}({\bf k})$ and $\mathbf{b}({\bf k})$ by the external forces $\mathbf{F}_{u}$ and $\mathbf{F}_{b}$; $D^{u}(\bold{k})$ and $D^{b}(\bold{k})$ are the dissipation rates of  $\mathbf{u}({\bf k})$ and $\mathbf{b}({\bf k})$, respectively; $T^{uu}\mathbf{(k)}$ and $T^{ub}\mathbf{(k)}$ are the modal energy transfers to mode $\mathbf{u}(\bold{k})$ from \textbf{u} and \textbf{b} fields, respectively; and $T^{bb}\mathbf{(k)}$ and $T^{bu}\mathbf{(k)}$ are the modal energy transfers to mode $\mathbf{b}(\bold{k})$ from \textbf{b} and \textbf{u} fields, respectively. Note that $T^{XY}$'s  occur due to the nonlinear interactions.
\begin{figure}
	\begin{center}
		\includegraphics[scale = 0.40]{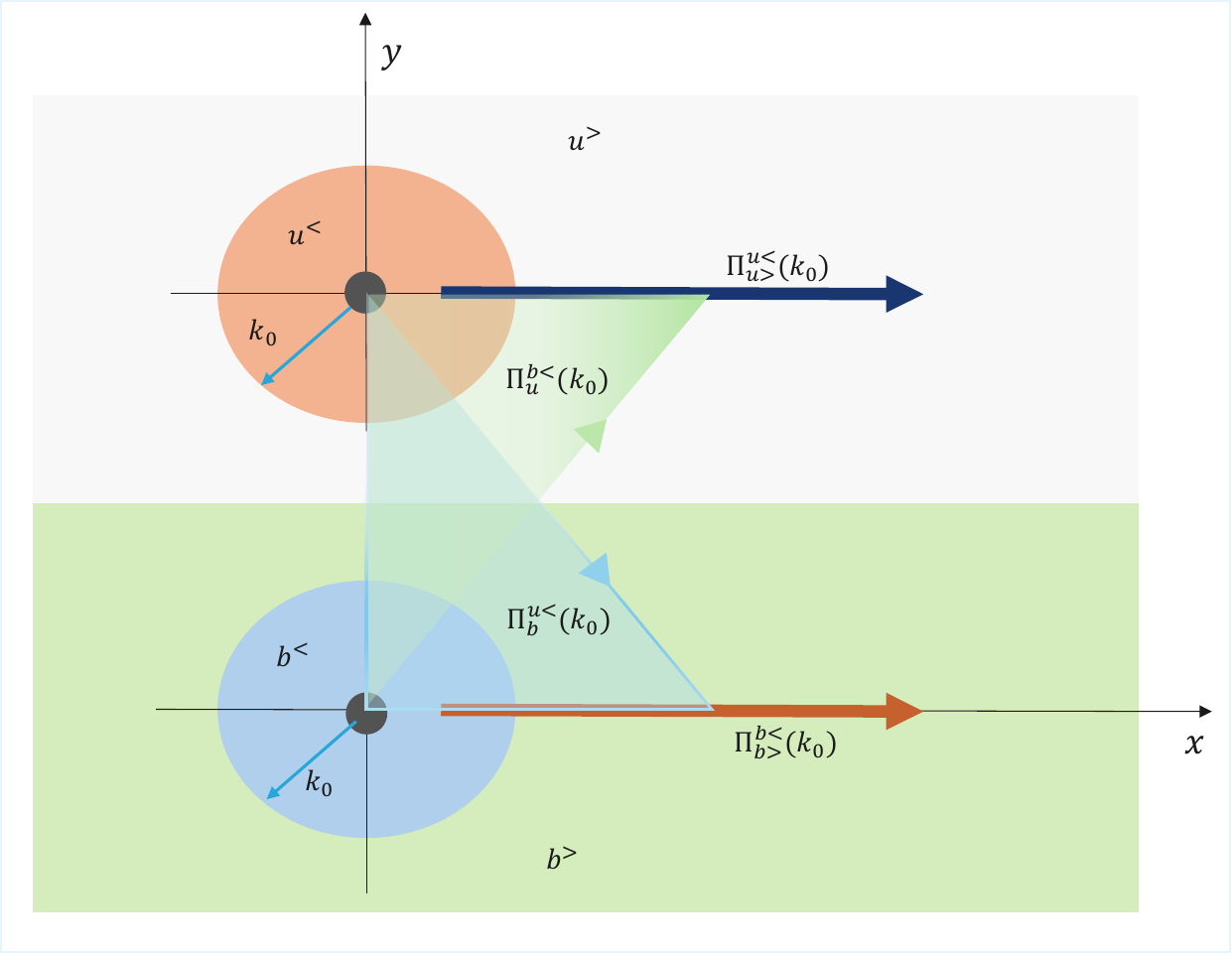}
	\end{center}
	\vspace*{0pt}
	\caption{Various energy fluxes of MHD turbulence. The definitions are given in Eqs.~(\ref{eq:flux_ug_ul}-\ref{eq:flux_u_bl}).}
	\label{fig:flux_transfer}
\end{figure}

The above four nonlinear terms yield four distinct energy fluxes for a wavenumber sphere of radius $k_0$~\cite{Verma:book:ET,Verma:JPA2022}:
\begin{align}
\Pi^{u<}_{u>}(k_0) &= -\sum_{|\mathbf{k}| \le k_0} T^{uu}(\mathbf{k}),
\label{eq:flux_ug_ul} \\
\Pi^{b<}_{b>}(k_0) &= -\sum_{|\mathbf{k}| \le k_0} T^{bb}(\mathbf{k}),
\label{eq:flux_bg_bl} \\
\Pi^{u<}_{b}(k_0) &= -\sum_{|\mathbf{k}| \le k_0} T^{ub}(\mathbf{k}),
\label{eq:flux_b_ul} \\
\Pi^{b<}_{u}(k_0) &= -\sum_{|\mathbf{k}| \le k_0} T^{bu}(\mathbf{k}),
\label{eq:flux_u_bl}
\end{align}
where $\Pi^{X<}_{X>}(k_0)$ denotes the net energy transfer from the modes $\mathbf{X(p)}$ within the wavenumber sphere of radius $k_0$ to the modes $\mathbf{X(k)}$ outside it, whereas $\Pi^{X<}_{Y}(k_0)$ denotes the energy transfer from the modes $\mathbf{X(p)}$ within the wavenumber sphere of radius $k_0$ to the modes $\mathbf{Y(k)}$ everywhere.  See Fig.~\ref{fig:flux_transfer} for an illustration. In this paper, we compute the respective fluxes using Eqs.~(\ref{eq:flux_ug_ul},\ref{eq:flux_bg_bl},\ref{eq:flux_b_ul},\ref{eq:flux_u_bl}). We remark that the above computations are more efficient than those computed using the mode-to-mode energy transfers~\cite{Verma:PR2004,Verma:JPA2022}. We also compute the perpendicular energy spectrum, $E_{T_\perp}(k_\perp)$,  using
\bea
E_{T_\perp}(k_\perp) &=& \frac{1}{2} \sum_{k_{\perp}-1\leq k_{\perp} ^ \prime < k_{\perp} }\lvert \mathbf{u_{\perp}(k^\prime)}\rvert^2 + \lvert \mathbf{b_{\perp}(k^\prime)}\rvert^2,
\label{eq:perp_spectrum} 
\eea
where $\mathbf{u_{\perp}(k)}$ and $\mathbf{b_{\perp}(k)}$ are, respectively, the components of $\mathbf{u(k)}$ and $\mathbf{b(k)}$ normal to the mean magnetic field $\mathbf{B_0}$.

For the Els\"{a}sser variables, the energy fluxes are~\cite{Verma:MHD_PRF,Verma:book:ET}:
\be
\Pi^\pm(k_0) = - \sum_{|\mathbf{k}| \le k_0} T^{\pm}(\mathbf{k}),
\ee
where $T^\pm({\bf k})$ are the modal energy transfer terms for ${\bf z}^\pm$~\cite{Verma:MHD_PRF}.  Note that the total energy flux is
\bea
\Pi_T(k_0) & = & \Pi^{u<}_{u>}(k_0)
+ \Pi^{u<}_{u>}(k_0)
+\Pi^{u<}_{u>}(k_0)
+\Pi^{u<}_{u>}(k_0) \nonumber \\
& = & \frac{1}{2} [\Pi^+(k_0)+\Pi^-(k_0)],
\eea 
whereas the cross helicity flux is
\bea
\Pi_{H_c}(k_0) & = &  
 \frac{1}{4} [\Pi^+(k_0)-\Pi^-(k_0)].
\eea 
In this paper, we will present the energy fluxes for \textbf{u}, \textbf{b}, the total energy, and the total cross helicity.

\renewcommand{\arraystretch}{1.5}
\begin{table*}
\caption{\label{tab:table1}Simulation parameters: 2D Runs 1-3 and the 3D Runs 4-6 are with $B_0=0$, and 2D Runs 7-8 are with $B_0=1$ and 3, respectively. The table lists the grid size, kinematic viscosity~($\nu$), Re, energy injection ratio  ($\epsilon^{H_c}_\mathrm{inj}/\epsilon^{T}_\mathrm{inj}$), time averaged $\la E_u / E_b \ra$, normalized cross helicity $\sigma_c$, Kolmogorov constant for MHD turbulence ($K_\mathrm{KoM}$). We also list the averaged $\la E_u(k) \ra / \la E_b(k)\ra$,  $\sigma_c(k) = 2\la H_c(k)\ra /\la E(k) \ra$, and $\la \Pi_{H_c}(k) \ra / \la \Pi_{T}(k) \ra$ in the inertial range.  For all the runs, $\eta = \nu$.}
\begin{tabular}{|>{\centering\arraybackslash}p{0.5cm}|>{\centering\arraybackslash}p{1cm}|>{\centering\arraybackslash}p{1cm}|>{\centering\arraybackslash}p{2cm}|>{\centering\arraybackslash}p{1.2cm}|>{\centering\arraybackslash}p{0.8cm}|>
{\centering\arraybackslash}p{1.2cm}|>
{\centering\arraybackslash}p{1.5cm}|>
{\centering\arraybackslash}p{1.6cm}|>
{\centering\arraybackslash}p{2.7cm}|>
{\centering\arraybackslash}p{1.4cm}|>
{\centering\arraybackslash}p{1.2cm}|>
{\centering\arraybackslash}p{1.2cm}} %
\toprule
 & $\mathbf{B_0}$ & Grid Size & $\nu = \eta$ & $\mathrm{Re}$ & $\frac{\epsilon^{H_c}_\mathrm{inj}}{\epsilon^{E}_\mathrm{inj}}$ & $\sigma_c$ & $\la \frac{ E_u}{ E_b}\ra$ & $ \frac{\la E_u(k) \ra}{\la E_b(k)\ra}$ &  $\sigma_c(k) = \frac{2\la H_c(k)\ra}{\la E(k)\ra}$ & $ \frac{\la \Pi_{H_c}(k)\ra}{ \la \Pi_T(k) \ra}$ & $K_\mathrm{KoM}$ \\ 
\toprule
$1$ & \multirow{6}{*}{$0$} & \multirow{2}{*}{$8192^2$} & $4.55 \times 10^{-5}$ & $96660$& $0$ & 0.03 & $0.75$ & $0.7$  & $0.0031$ & $-0.001$ & $4.6$ \\ 
\cline{1-1} \cline{4-12}
$2$ &  &   & $6.6 \times 10^{-5}$ & $104700$& $1/4$ & $0.65$ & $0.77$ & $0.84$  & $0.61$ & $0.225$ & $5.5$ \\ 
\cline{1-1} \cline{4-12}
$3$  & & & $8 \times 10^{-5}$ & $133500$& $1/3$ & 0.82 & $0.93$ & $0.86$  & $0.65$ & $0.25$ & $6.3$ \\
\cline{1-1} \cline{3-12}
$4$ & & \multirow{2}{*}{$1536^3$} & $1.5 \times 10^{-4}$ & $26780$ & $0$ & $0.018$ & $0.57$ & $0.48$ & $0.007$ & $-0.002$ & $2.01$ \\ 
\cline{1-1} \cline{4-12}
$5$ & &  & $2.15 \times 10^{-4}$  & $25130$ & $1/4$ & $0.703$ & $0.92$ & $0.61$  & $0.65$ & $0.23$ & $2.5$ \\ 
\cline{1-1} \cline{4-12}
$6$ & & & $2.4 \times 10^{-4}$ & $32369$ & $1/3$ & $0.84$ & $0.78$ & $0.69$ & $0.81$ & $0.31$ & $4.1$ \\ 
\hline
$7$ & $1$ & \multirow{2}{*}{$8192^2$} & $4.55 \times 10^{-5}$  & $98206$ & $0$ & $0.001$ & $1.28$ & $0.88$  & $-0.05$ & $-0.017$ & $5.02$ \\ 
\cline{1-2} \cline{4-12}
$8$ & $3$ &  & $1.47 \times 10^{-5}$ & $522040$ & $0$ & $0.07$ & $1.1$ & $1.02$ & $0.07$ & $0.0052$ & $8.31$ \\ 
\hline
\end{tabular}
\label{tab:table}
\end{table*}

Additionally, \citet{Politano:GRL1998,Politano:PRE1998} generalized Kolmogorov's structure-function derivation to isotropic MHD turbulence  (${\bf B}_0=0$) and derived the following  \textit{exact relations} for the third-order structure functions for the total energy and cross helicity  in the limit of vanishing viscosity and magnetic diffusivity:
\bea
S^T_3(l) &=& \left\langle |\Delta \mathbf{v}|^2 (\Delta \mathbf{v \cdot \hat{l}}) \right\rangle 
+ \left\langle |\Delta \mathbf{b}|^2 (\Delta \mathbf{v \cdot \hat{l}}) \right\rangle \nonumber \\
&& - 2\left\langle (\Delta \mathbf{v} \cdot \Delta \mathbf{b}) (\Delta \mathbf{b} \cdot \hat{\mathbf{l}}) \right\rangle
=- \frac{4}{d} \epsilon^T l, \label{eq:S3_T} \\
S^{H_C}_3(l) &=& -\left\langle |\Delta \mathbf{b}|^2 (\Delta \mathbf{b \cdot \hat{l}}) \right\rangle 
- \left\langle |\Delta \mathbf{v}|^2 (\Delta \mathbf{b \cdot \hat{l}}) \right\rangle \nonumber \\
&& +2\left\langle (\Delta \mathbf{v} \cdot \Delta \mathbf{b}) (\Delta \mathbf{v \cdot \hat{l}}) \right\rangle
=- \frac{8}{d} \epsilon^{H_C} l, \label{eq:S3_HC}
\eea
respectively, where $ \Delta \mathbf{u} = \mathbf{u(x+l) - u(x)} $ and $ \Delta \mathbf{b} = \mathbf{b(x+l) - b(x)} $; $d$ is the space dimension; and $\epsilon^T$ and $\epsilon^{H_C}$ are the dissipation rates of the total energy and cross helicity, respectively. The above structure functions indicate that the second-order correlation functions of the field variables are expected to scale as $l^{2/3}$, which leads to $k^{-5/3}$ energy spectrum. Thus, the structure function predictions by \citet{Politano:GRL1998,Politano:PRE1998} favor Kolmogorov scaling rather than IK scaling. We will compute $S^T_3(l)$ and $S_3^{H_C}(l)$ using numerical data and verify the above exact relations [Eqs.~(\ref{eq:S3_T}, \ref{eq:S3_HC})]. Unfortunately, isotropic $S_3(l)$ is not defined for anisotropic the case ($\mathbf{B_0} \ne 0$). Hence, we do not compute $S_3(l)$  for the anisotropic case.

\section{Numerical details}
\label{sec:numerical_details}
Contrasting Kolmogorov and IK scalings necessitates high-resolution simulations, which are computationally expensive in 3D. Fortunately, \textit{absolute~equilibrium theory} for MHD turbulence~\cite{Frisch:JFM1975,Kraichnan:ROPP1980,Biskamp:book:MHDTurbulence} predicts that the spectra and fluxes of the total energy are similar in both 2D and 3D.  Consequently, researchers have employed reduced-cost 2D MHD turbulence simulations to investigate the scaling~\cite{Biskamp:PFB1989,Verma:JGR1996DNS}. However, according to \textit{antidynamo theorem}~\cite{Cowling:book}, in 2D, the magnetic field vanishes when only \textbf{u} field is forced. Therefore,  we force both \textbf{u} and \textbf{b} fields (or ${\bf z}^\pm$) to obtain a steady state in 2D MHD.  We performed our simulations on \textit{Frontier} and IIT Kanpur HPC systems using \textit{Aithon}, a CUDA C++ code.  
Aithon is roughly 3000 times faster on an A100 GPU than on a single core of an AMD EPYC 7742 processor. Additionally, we use GPUs to speed up the computations of the energy spectra, fluxes, and structure functions.

We employ pseudospectral method to simulate MHD turbulence in a  $(2\pi)^2$ domain with a $8192^2$ grid and in a $(2\pi)^3$ domain with a $1536^3$ grid. We also simulate anisotropic MHD with $\mathbf{B_0} = \hat{z}$ and $3\hat{z}$ on $8192^2$ grid. The simulation details are listed in Table~\ref{tab:table}. Energy is injected in the wavenumber band (2,3) at a fixed rate of $\epsilon^-_\mathrm{inj}=0.1$ for ${\bf z}^-$  and at a varying rates of  $\epsilon^+_\mathrm{inj} = 0.1$, 0.3, and 0.5 for ${\bf z}^+$, which yield $\epsilon^{H_c}_\mathrm{inj}/ \epsilon^T_\mathrm{inj}=0$, 1/4, and 1/3, respectively~\cite{Verma:MHD_PRF}. The runs with $\epsilon^+_\mathrm{inj} \ne \epsilon^-_\mathrm{inj}$ induce Alfv\'{e}nicity (nonzero $\sigma_c$, as shown in Table~\ref{tab:table}) that helps in contrasting the IK and Kolmogorov scalings.
In addition, the injection rates of magnetic helicity ($\epsilon_\mathrm{inj}^{H_M}$) in 3D and of the mean-square vector potential ($\epsilon_\mathrm{inj}^{A}$) in 2D are negligible. For the anisotropic cases, we set $\epsilon_\mathrm{inj}^{H_c} = 0$.

\begin{figure*}
	\begin{center}
		\includegraphics[scale = 0.65]{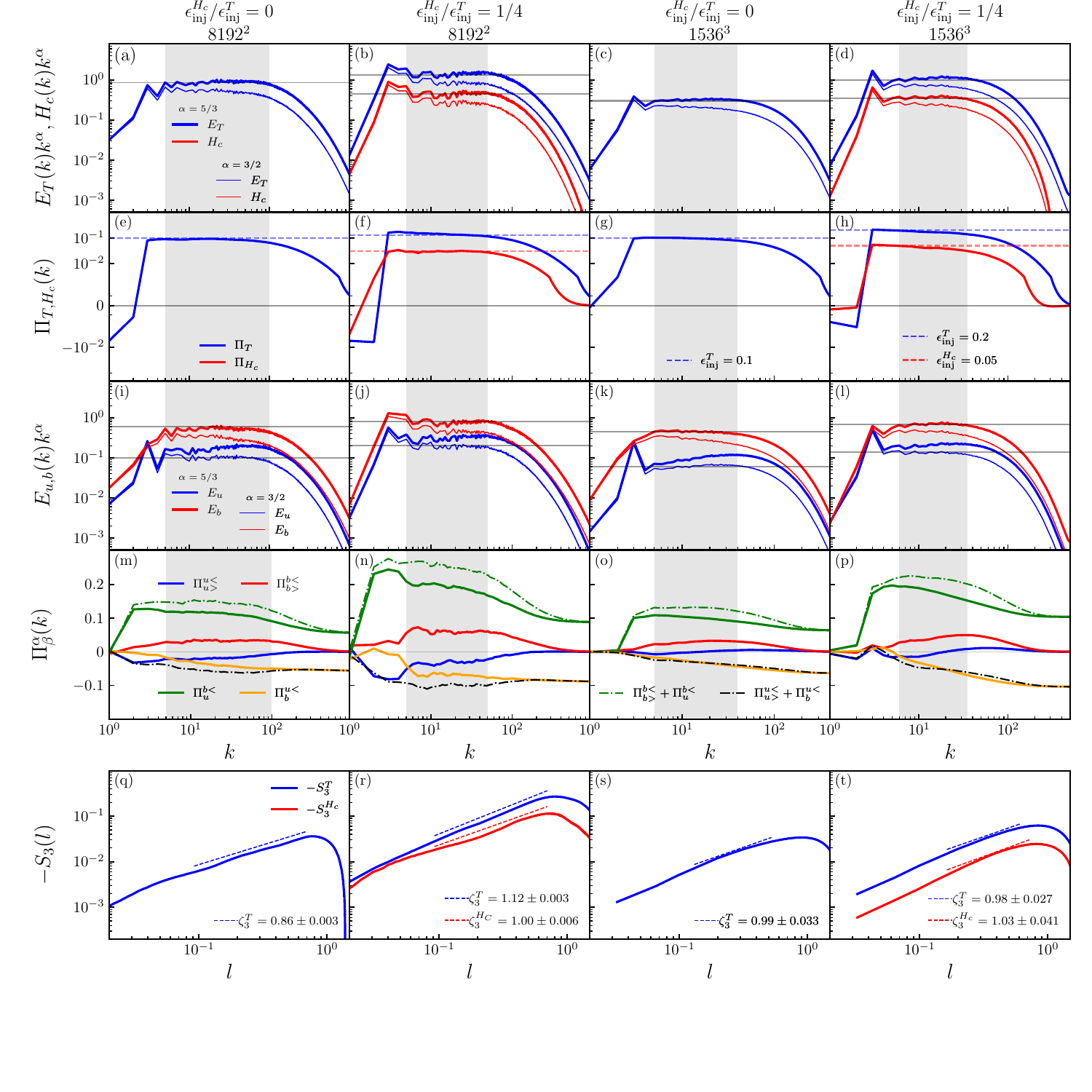}
	\end{center}
	\vspace*{0pt}
	\caption{Numerical results of isotropic MHD turbulence ($B_0=0$) simulations on $8192^2$ (2D) and $1536^3$ (3D) grids for energy injection ratios $\epsilon^{H_C}_\mathrm{inj}/\epsilon^T_\mathrm{inj} = 0$ and 1/4. Panels (a–d) show the normalized energy spectra $E_T(k),H_c(k)\,k^\alpha$ with $\alpha = 5/3$ or $3/2$. Panels (e–h) display the corresponding energy fluxes $\Pi_{T,H_c}(k)$ that match with the respective injection rates (dashed lines). Panels (i–l) present the normalized energy spectra  $E_u(k) k^\alpha$ and $E_b(k) k^\alpha$. { Panels (m–p) show energy fluxes   $\Pi^{u<}_{u>}(k)$ (blue), $\Pi^{b<}_{b>}(k)$ (red), $\Pi^{b<}_u(k)$ (green),  $\Pi^{u<}_b(k)$ (orange), $\Pi^{u<}_{u>}(k)+\Pi^{u<}_{b}(k)$ (black chained), and $\Pi^{b<}_{b>}(k)+\Pi^{b<}_{u}(k)$ (green chained).} Panels (q–t) depict $-S_3^{T,H_c}(l) \propto l$ for the total energy and cross helicity.}
	\label{fig:figure}
\end{figure*}
\begin{figure}
	\begin{center}
		\includegraphics[scale = 0.55]{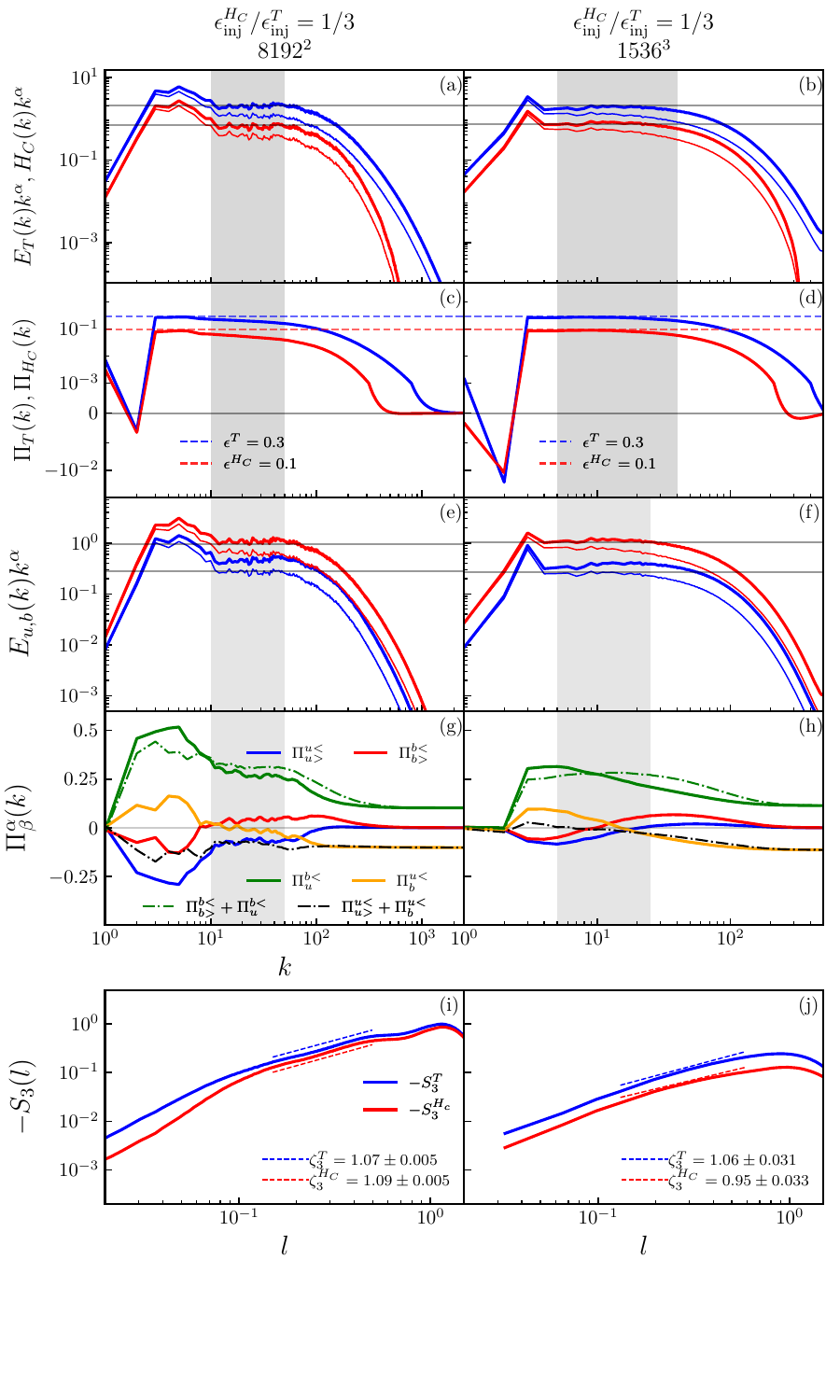}
	\end{center}
	\vspace*{0pt}
	\caption{Numerical results of isotropic MHD turbulence ($B_0=0$)  simulations on $8192^2$ (2D) and $1536^3$ (3D) grids for energy injection ratio of $\epsilon^{H_C}_\mathrm{inj}/\epsilon^T_\mathrm{inj} = 1/3$.  Refer to Fig.~\ref{fig:figure} for the description of the plots.}
	\label{fig:Ek_Pik_1by3}
\end{figure}

For all the runs, the viscosity and magnetic diffusivity are  equal, and the maximum kinetic and magnetic Reynolds numbers for 2D and 3D are 522040 and 32369, respectively.   We employ hypo-viscosity and hypo-diffusivity for the 2D runs to suppress the growth of $E_b(k)$  due to the inverse cascade of $A^2$.  With Kolmogorov length as $l_d = [\nu/(\epsilon^+_\mathrm{inj})^3]^{1/4} $,  $k_\mathrm{max} l_d$ for our runs range from 1.84 to 4.0. Hence, all our runs are fully resolved.  Averaged hydrodynamic entropies~\cite{Verma:PRF2022,Verma:PRE2024}
for the six isotropic runs are 5.19, 4.38, 4.01, 4.7, 4.3, and 4.2, respectively, which are near those for hydrodynamic turbulence~\cite{Verma:PRE2024}. Note that the hydrodynamic entropy decreases with the increase of $\sigma_c$ for both 2D and 3D, implying that the flows get more ordered with the injection of cross helicity. We conduct our simulations for a significantly long time, which yields numerous temporally separated dataframes, allowing for effective averaging of structure functions, energy spectra, and fluxes.  See Table~\ref{tab:table} and the companion paper \cite{Verma:MHD_PRF} for details.

\section{Numerical results}
\label{sec:results}

In this section, we present our numerical results for both isotropic and anisotropic cases. At the end of the section, we compare our results with those from solar wind and field theories.

\subsection{Isotropic Case}

\begin{figure*}
	\begin{center}
		\includegraphics[scale = 0.55]{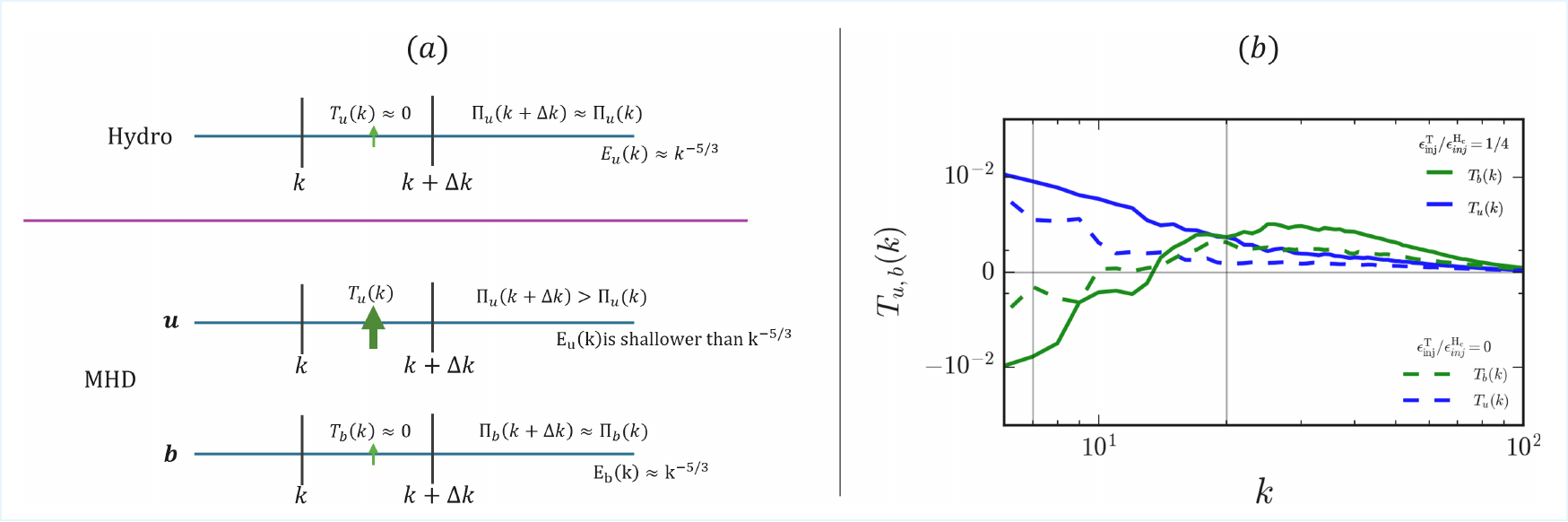}
	\end{center}
	\vspace*{0pt}
	\caption{(a) Schematic diagrams of energy transfer function $T(k)$ in Hydrodynamic and MHD turbulence.   
    In hydrodynamic turbulence $T_u(k) = d\Pi_u(k)/dk \approx 0$, implying constancy of energy flux and $k^{-5/3}$ energy spectrum.     Similarly, for our MHD turbulence, $T_b(k) \approx 0$, leading to $E_b(k) \sim k^{-5/3}$ spectrum. But, $T_u(k) > 0$, leading to $E_u(k)$ shallower than $k^{-5/3}$.  See Eqs.~(\ref{eq:T_b},\ref{eq:T_u}). 
    (b) For the 3D simulations on $1536^3$ grid with $\epsilon^{H_C}_\mathrm{inj}/\epsilon^T_\mathrm{inj} = 1/4$ (solid curves) and $0$ (dashed curves) (third and fourth columns of Fig.~\ref{fig:figure}), the 
    numerically computed $T_{b,u}(k)$. We observe that $T_b(k)$ is nearly 0 on an average, but $T_u(k)$ is strongly positive.
    }
	\label{fig:Tk_figure}
\end{figure*}

Now we present our numerical results, starting with the energy spectra and fluxes. Figure~\ref{fig:figure} presents the spectra, fluxes, and structure functions for ${\epsilon^{H_c}_\mathrm{inj}}/{\epsilon^{E}_\mathrm{inj}}=0$ and 1/4, whereas Fig.~\ref{fig:Ek_Pik_1by3} presents these quantitites for
${\epsilon^{H_c}_\mathrm{inj}}/{\epsilon^{E}_\mathrm{inj}}=1/3$. Figure~\ref{fig:figure}(a,b,c,d) and Fig.~\ref{fig:Ek_Pik_1by3}(a,b) (first rows) illustrate the normalized spectra for the total energy ($E_T(k) k^\alpha$) and the cross helicity  ($H_c(k) k^\alpha$)  with $\alpha = 5/3$ and 3/2; the first two columns are for $8192^2$ grid and the next two columns are for $1536^3$ grid.  As shown in the figure, Kolmogorov’s spectrum provides a better fit to the data than IK spectrum, but the contrast is not very sharp. The second rows of Fig.~\ref{fig:figure} and Fig.~\ref{fig:Ek_Pik_1by3} exhibits constant inertial-range fluxes for the total energy and cross helicity; these fluxes match with the respective injection rates (shown as dashed lines). Kolmogorov's constants $K_\mathrm{KoM}$ computed using Eq.~(\ref{eq:Kolm}), listed in Table~\ref{tab:table}, are in general agreement with past numerical and field-theoretic works  \cite{Beresnyak:PRL2011,Verma:PR2004,Verma:NC2025} (within factor of 2). An increase of $K_\mathrm{KoM}$ with $\sigma_c$ indicates a relative suppression of the total energy flux with cross helicity and mean magnetic field. A more effective strategy is to test whether $\Pi_{H_c}(k) \approx 0$, as in the IK scaling, or whether $\Pi_{H_c}(k)$ follows the trend of $\sigma_c$, as in Kolmogorov scaling [Eq.~(\ref{eq:Kolm_pm})].  As shown in Fig.~\ref{fig:figure}(e-h) and Fig.~\ref{fig:Ek_Pik_1by3}(c-d), in the inertial range, $\Pi_{H_c}(k) \approx \epsilon^{H_c}_\mathrm{inj}$, as predicted by Kolmogorov scaling.  \citet{Verma:JGR1996DNS} arrived at similar conclusions, but their grid resolutions were too coarse.

Next, we report $E_u(k)$ and $E_b(k)$. The third rows of Figs.~\ref{fig:figure} and \ref{fig:Ek_Pik_1by3} show that $E_b(k) \sim k^{-5/3}$ and $E_u(k) \sim k^{-3/2}$, similar to many past results~\cite{Podesta:ApJ2007,Alexakis:PRL2013,Jiang:JFM2023}. {These observations appear to contradict both Kolmogorov or IK scaling, but they can be easily explained in the framework of \textit{variable energy flux}~\cite{Verma:book:ET,Verma:JPA2022,Verma:Fluid2021} using the energy transfers from the magnetic field to the velocity field.} The fourth rows of Fig.~\ref{fig:figure} and Fig.~\ref{fig:Ek_Pik_1by3} present the energy fluxes for the sphere of radius $k$: $\Pi^{u<}_{u>}(k), \Pi^{b<}_{b>}(k), \Pi^{u<}_{b}(k), \Pi^{b<}_{u}(k)$, defined in Eqs.~(\ref{eq:flux_ug_ul}-\ref{eq:flux_u_bl}) \cite{Verma:MHD_PRF,Dar:PD2001,Verma:PR2004}. 
Note that these fluxes remain constant over time during the steady state. Among them, the kinetic energy flux, $\Pi^{u<}_{u>}(k)\approx 0$  (blue curves) in 3D, but $\Pi^{u<}_{u>}(k)<0$ in 2D, which is reminiscent of the inverse energy cascade in 2D hydrodynamic turbulence.  The magnetic energy flux $\Pi^{b<}_{b>}(k)$  (red curves) is also relatively small compared to the magnetic-to-kinetic energy flux [$\Pi^{b<}_{u}(k)$ (green curves)] and kinetic-to-magnetic energy flux [$\Pi^{u<}_{b}(k)$ (orange curves)]. 

Interestingly, the energy fluxes provide clues why $E_u(k)$ appears as $k^{-3/2}$. The arguments are as follows:
\begin{enumerate}
   \item In hydrodynamic turbulence, the constant inertial-range energy flux leads to vanishing of the  nonlinear energy transfer, i.e., $T_u(k) = -d \Pi_u(k)/dk \approx 0$. Hence, a  thin wavenumber shell of radius $k$ does not receive or lose energy per unit time. Therefore, the net flux $k u_k^3 \approx $ const., employing that $u_k \sim k^{-1/3}$ or $E_u(k) \propto k^{-5/3}$ (see Fig.~\ref{fig:Tk_figure}(a)).

    \item For the forcing employed in our paper, as shown in  Figs.~\ref{fig:figure} and \ref{fig:Ek_Pik_1by3}, the net energy flux from a magnetic sphere  turbulence, $\Pi^{b<}_{b>}(k) + \Pi^{b<}_{u}(k) \approx $ constant in the inertial range (green chained curves). Therefore [Eqs.~(\ref{eq:Ek_b},\ref{eq:flux_bg_bl},\ref{eq:flux_u_bl}) and Fig.~\ref{fig:Tk_figure}],
    \be
    T_b(k) = -\frac{d}{dk} [\Pi^{b<}_{b>}(k) + \Pi^{b<}_{u}(k)] \approx 0,
    \label{eq:T_b}
    \ee
    similar to that in hydrodynamic turbulence. Hence, the corresponding energy spectrum $E_b(k) \propto k^{-5/3}$.  

    \item As shown in Figs.~\ref{fig:figure} and \ref{fig:Ek_Pik_1by3}, the velocity field \textbf{u} gains energy.  More importantly,  Fig.~\ref{fig:Tk_figure} and Eqs.~(\ref{eq:Ek_u},\ref{eq:flux_ug_ul},\ref{eq:flux_b_ul})  show that
    \be
    T_u(k) = -\frac{d}{dk} [\Pi^{u<}_{u>}(k) + \Pi^{u<}_{b}] > 0.
    \label{eq:T_u}
    \ee
    Therefore, for the velocity field, a  thin wavenumber shell of radius $k$ receives  energy per unit time. Therefore, the normalized flux increases with $k$. Consequently, $E_u(k)$ is shallower than $k^{-5/3}$ energy spectrum. The shallow $E_u(k)$ gives an impression of $k^{-3/2}$, but this is not a universal feature. For a different forcing, \citet{Verma:Fluid2021} showed that  $E_b(k)$ is shallower than $E_u(k)$ due to energy transfer from \textbf{u} to \textbf{b}. Note that the spectral exponents 5/3 and 3/2 for $E_u(k)$ and $E_b(k)$ are quite close, and they are hard to distinguish.
\end{enumerate}
The above energetics-based arguments are qualitative.  In the future, we will better quantify $T_u(k)$ and $T_b(k)$ by averaging over many more frames, and study the variance of spectral indices from 5/3. Additionally, the above energy transfers are consistent with those reported earlier~\cite{Debliquy:PP2005,Dar:PD2001,Verma:PP2005}. An exception to the above general pattern is the 2D run with $\epsilon^{H_C}_\mathrm{inj}/\epsilon^T_\mathrm{inj} = 1/4, 1/3$,  for which $\Pi^{u<}_{u>}(k) + \Pi^{u<}_{b}(k)$ is nearly constant. This issue needs further investigation in the light of nonlocal energy transfers observed in 2D turbulence~\cite{Dar:PD2001,Gupta:PRE2019}. 

The above results indicate that the energy fluxes of kinetic and magnetic energies vary with $k$, in contrast to the constant energy fluxes of the total energy, cross helicity, and $E^\pm$, which are inviscid invariants. Hence, $E_u(k)$ and $E_b(k)$ are unlikely to yield a correct spectral exponent, but the total energy, cross helicity, and $E^\pm$ do because they are inviscid invariants~\cite{Verma:book:ET,Verma:JPA2022}.


The fifth rows of Fig.~\ref{fig:figure} and \ref{fig:Ek_Pik_1by3} exhibit the averaged structure functions $S^{T,H_c}_3(l)$  for the total energy and cross helicity (when $B_0=0$).  We observe that $S^{T,H_c}_3(l) \propto l $,  consistent with Kolmogorov scaling~\cite{Politano:GRL1998,Politano:PRE1998}. These results are consistent with several past works~\cite{Biskamp:PP2000,Biskamp:PP2001,Sorriso:PP2002,Mininni:PRE2009a,Jiang:JFM2023}. Recently, \citet{Verma:MHD_PRF} analyzed the intermittency exponents for MHD turbulence and found them to be closer to Komogorov scaling than IK scaling (generalization of She-Leveque theory to MHD~\cite{She:PRL1994}); these results are consistent with the predictions of  \citet{Politano:GRL1998,Politano:PRE1998} and others. In this paper, we do not discuss  the dynamic alignment \cite{Boldyrev:PRL2006}, which is typically studied for anisotropic MHD (${\bf B}_0 \ne 0$).

\subsection{Anisotropic Cases}
\label{sec:result_anisotropic_cases}

\begin{figure}
	\begin{center}
		\includegraphics[scale = 0.55]{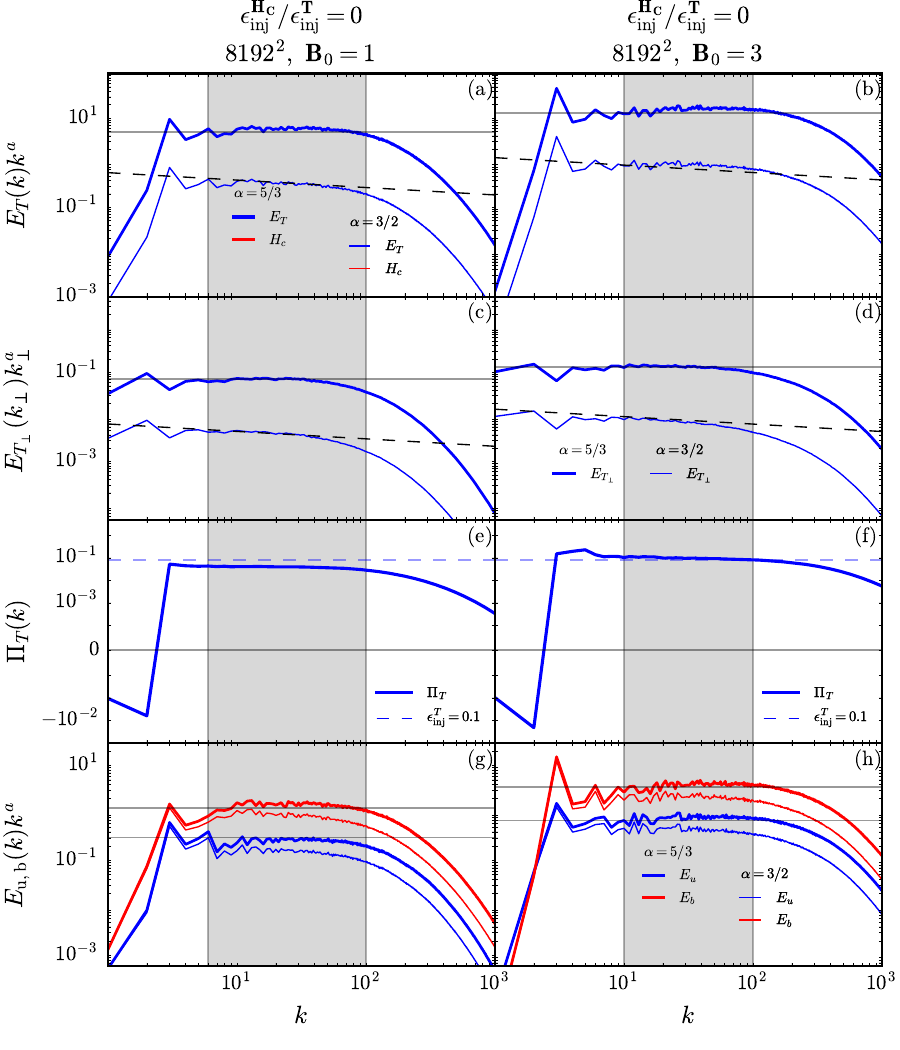}
	\end{center}
	\vspace*{0pt}
	\caption{Numerical results of MHD turbulence simulations on $8192^2$ (2D) grid for energy injection ratios $\epsilon^{H_C}_\mathrm{inj}/\epsilon^T_\mathrm{inj} = 0$ with $\mathbf{B_0} = 1$ (1st coloum) and $3$ (2nd coloum). Panels (a,b) show the normalized energy spectra $E_T(k)\,k^{a}$ with $a = 5/3$ or $3/2$. Panels (c–d) display the corresponding perpendicular energy spectra $E_{T_\perp}(k_\perp)\,k_\perp^{a}$ with $a = 5/3$ or $3/2$. Panels (e,f) shows total energy fluxe $\Pi_{T}(k)$ that match with the respective injection rates (dashed lines). Panels (g,h) present the normalized energy spectra  $E_u(k) k^\alpha$ and $E_b(k) k^\alpha$.}
	\label{fig:anisotropy}
\end{figure}

In this subsection, we present results from anisotropic MHD simulations with a uniform mean magnetic field $\mathbf{B}_0 = B_0 \hat{z}$, with $B_0 = 1$ and $B_0 = 3$. Figure~\ref{fig:anisotropy} shows the simulation results at a resolution of $8192^2$ with zero cross-helicity injection, $\epsilon^{\mathrm{H_c}}_{\mathrm{inj}} / \epsilon^{\mathrm{T}}_{\mathrm{inj}} = 0$. The first column corresponds to $B_0 = 1$, while the second column corresponds to $B_0 = 3$.

In Fig.~\ref{fig:anisotropy}(a,b), the total energy spectrum compensated by $k^{5/3}$ (thick curve) is constant in the inertial range, indicating scaling consistent with Kolmogorov phenomenology~\cite{Biskamp:book:MHDTurbulence,Verma:PR2004,Dar:PD2001}. In contrast, the compensated spectrum $E_T(k)k^{3/2}$ (thin curve)  follows a $k^{-1/6}$ trend (dashed line), which arises from the difference between the $k^{-5/3}$ and $k^{-3/2}$ scaling. These observations are consistent with Kolmogorov scaling, as in the isotropic case.

Figure~\ref{fig:anisotropy}(c,d) shows the perpendicular energy spectrum computed using Eq.~(\ref{eq:perp_spectrum}). The spectrum exhibits a  $k^{-5/3}$ scaling (thick line) in the inertial range, consistent with \citet{Goldreich:ApJ1995} predictions. Figure~\ref{fig:anisotropy}(e,f) presents the total energy flux. Within the inertial range (shaded region), the flux for the total energy is constant and matches with the energy injection rate, indicated by the dashed line. An inverse energy cascade at large scales, characteristic of two-dimensional turbulent flows, is also evident. The compensated kinetic (blue curve) and magnetic (red curve) energy spectra  shown in Fig.~\ref{fig:anisotropy}(g,h) support Kolmogorov-like scaling. When compensated by $k^{3/2}$, neither $E_u(k)k^{3/2}$ nor  $E_b(k)k^{3/2}$  flattens, further excluding Kraichnan-Iroshnikov scaling for the anisotropic MHD turbulence. 

For the anisotropic case, the kinetic and magnetic energy spectra follow $k^{-5/3}$ scaling. In contrast, in the isotropic case, the kinetic energy spectrum follows $k^{-3/2}$ scaling, while the magnetic energy spectrum follows $k^{-5/3}$ scaling.  The above feature appears to follow because of the propagating Alfv\'{e}n waves, which may suppress energy exchange between \textbf{u} and \textbf{b}.

The anisotropic results presented here correspond to mean magnetic field strengths $B_0 = 1$ and $B_0 = 3$. Simulations with stronger mean magnetic field, together with a more detailed analysis of their spectral properties, will be reported in future work.

\subsection{Comparison with Solar Wind Observations and Field Theory}
Next, we compare the above numerical results with  solar wind observations. Researchers have computed the energy spectra $E_u(k), E_b(k)$, and $E^\pm(k)$ of the solar wind using Taylor’s frozen-in hypothesis~\cite{Verma:PP2022}. Some researchers~\cite{Matthaeus:JGR1982rugged,Tu:SSR1995,Goldstein:ARAA1995,Matthaeus:SSR2011,Petrosyan:SSR2010} support Kolmogorov scaling, but  others  \cite{Bourouaine:ApJ2020} support IK scaling. Additionally, \citet{Podesta:ApJ2007} reported Kolmogorov scaling for the magnetic field and IK scaling for the velocity field, as in our runs.  \citet{SorrisoValvo:PRL2007}'s structure-function estimates using spacecraft data are in reasonable agreement with Eqs.~(\ref{eq:S3_T}, \ref{eq:S3_HC}), consistent with Kolmogorov’s scaling.  Considering these divergences in the solar wind results, our numerical results are of immense value.  

Field theory computations also help validate MHD turbulence models.  Refer to the earlier discussions for the analytical works on anisotropic MHD turbulence \cite{Sridhar:ApJ1994,Goldreich:ApJ1995,Galtier:JPP2000}.  For isotropic MHD turbulence, renormalization group analysis and energy flux calculations mostly support Kolmogorov scaling \cite{Verma:PP1999,Verma:PRE2001,Verma:PR2004,Adzhemyan:book:RG,Verma:NC2025,Zhou:PR2010}, but there are variants. \citet{Verma:PP1999} showed that the renormalized mean magnetic field $B_0$ scales as $(\epsilon^T)^{1/3} k^{-1/3}$, substitution of which in Eq.~(\ref{eq:IK}) leads to $E_T(k) \sim (\epsilon^T)^{2/3} k^{-5/3} $. Thus, \citet{Verma:PP1999}  shows that Eq.~(\ref{eq:IK}) can lead to Kolmogorov scaling for strong MHD turbulence.

\section{Discussion and Conclusions}

Identifying the correct turbulence phenomenology for isotropic and anisotropic MHD turbulence---whether Kolmogorov scaling ($k^{-5/3}$) or Iroshnikov-Kraichnan scaling ($k^{-3/2}$)---remains uncertain.  In this paper, we address this issue using high-resolution numerical simulations. Since the spectral exponents $-5/3$ and $-3/2$ are close, we focus on the fluxes and structure functions of the total energy and cross helicity and convincingly establish Kolmogorov-like phenomenology for MHD turbulence. 

We also show that the spectral exponents of the velocity and magnetic fields are not accurate due to the energy exchanges between them. Instead, the total energy, cross helicity, and the energy of Els\"{a}sser variables should be used for spectral diagnostics because their inertial-range fluxes are constant~\cite{Verma:MHD_PRF,Verma:JPA2022}.   These results significantly enhance our ability to accurately model astrophysical turbulence, such as in the solar corona, dynamos, and the solar wind.

For anisotropic MHD turbulence with  mean magnetic fields ${\bf B}_0 = \hat{z}$ and $3\hat{z}$, our results demonstrate that the total energy spectrum, the perpendicular energy spectrum, and the total energy flux are all consistent with Kolmogorov-like phenomenology. Furthermore, the kinetic and magnetic energy spectra exhibit Kolmogorov scaling, which may be due to the suppression of energy exchange between \textbf{u} and \textbf{b} in anisotropic MHD turbulence.

At present, a reliable formulation for the third-order structure function $S_3(l)$ in anisotropic MHD turbulence is not available, which prevents us from carrying out an exact scaling analysis based on third-order statistics. Several important issues in anisotropic MHD turbulence still remain unresolved. A more comprehensive investigation of anisotropic MHD turbulence, including simulations with large $B_0$ and a detailed examination of associated spectral properties, is planned for the future.

\vspace{0.2cm}

Acknowledgments.-- The authors thank Melvyn Goldstein, Aaron Roberts, William Matthaeus, Riddhi Bandyopadhyay, Jayant Bhattacharjee, Alexander Schekochihin, Stephan Fauve, Alex Alexakis, Rodion Stepanov, Franck Plunian, Soumyadeep Chatterjee, Amit Agrawal, and Shashwat Nirgudkar for valuable suggestions at various stages of our investigation. This research used Frontier of the Oak Ridge Leadership Computing Facility (through Director’s Discretionary Program) at the Oak Ridge National Laboratory.    Simulations were also performed on Param Sanganak (IIT Kanpur), and HPC facility of Kotak School of Sustainability (KSS). Part of this work
was done in the Center for Turbulence Research, Stanford University, where MKV was a Visiting Senior Fellow.   Part of this work was supported
by  the J. C. Bose
Fellowship (SERB /PHY/2023488) and a project from Kotak School of Sustainability (DORA /DORA/2023508I).


%
\end{document}